\documentclass[notoc,12pt]{JHEP}

\usepackage{amsmath,amssymb,euscript,array,cite}

\input{epsf}

\usepackage{epsfig}

\def\N{{\cal N}}

\def\ttau{{\tilde\tau}}

\def\Tr{{\rm Tr}}

\def\det{{\rm det}}
\def\SU{\text{SU}}
\def\U{\text{U}}

\def\SL{\text{SL}}

\def\Dbarslash{\,\,{\raise.15ex\hbox{/}\mkern-12mu {\bar\D}}}
\def\Dslash{\,\,{\raise.15ex\hbox{/}\mkern-12mu \D}}
\def\delslash{\,\,{\raise.15ex\hbox{/}\mkern-9mu \partial}}
\def\delbarslash{\,\,{\raise.15ex\hbox{/}\mkern-9mu {\bar\partial}}}

\newcommand{\EQ}[1]{\begin{equation} #1 \end{equation}}

\newcommand{\SP}[1]{\begin{equation}\begin{split} #1 \end{split}\end{equation}}

\title{Massive Vacua of $\N=1^*$ Theory and $S$-duality from Matrix Models} 
\author{Nick Dorey, Timothy J.~Hollowood, S.~Prem~Kumar and Annamaria
Sinkovics\\
Department of Physics, University of Wales Swansea,
Swansea, SA2 8PP, UK\\
E-mail: {\tt t.hollowood@swan.ac.uk}, {\tt n.dorey@swan.ac.uk}
{\tt s.p.kumar@swan.ac.uk}}
\preprint{SWAT-352}
\abstract{In this note we show how Dijkgraaf and Vafa's hypothesis
relating the exact
superpotential of an $\N=1$ theory to a matrix model 
can be used to describe all the
massive vacua of the $\N=1^*$, or mass deformed $\N=4$, theory
including the Higgs vacuum. The matrix model computation of the
superpotential for each massive vacuum independently yields a modular
function of the associated effective coupling in that vacuum which
agrees with previously derived results up to a vacuum-independent
additive constant. The 
results in the different massive vacua can be related by the action of
$\SL(2,{\mathbb Z})$ on the $\N=4$ coupling, thus providing
evidence for modular invariance of the underlying $\N=4$ theory.}

\begin{document}

\section{Introduction}

According to Dijkgraaf and Vafa's (DV) proposal 
\cite{Dijkgraaf:2002fc,Dijkgraaf:2002vw,talk,Dijkgraaf:2002dh},
the effective superpotentials for a 
large class of $\N=1$ supersymmetric field theories at {\it finite\/} $N$ 
are computed from related large-$N$ matrix models. Some of the best tests
of this hypothesis have been for certain mass deformations of $\N=4$
supersymmetric gauge theory which preserve $\N=1$ supersymmetry---the 
so-called $\N=1^*$ theory. The latter theory is obtained by adding
masses for all three adjoint chiral superfields of the $\N=4$
theory. In addition, one can allow arbitrary polynomial deformations
of the superpotential involving one of the three adjoint
superfields. In this note, we will refer to this entire class of theories as
$\N=1^*$ theories.
These theories provide some of the best tests of the DV proposal,
since the resulting effective
superpotential depends in a non-trivial and characteristic 
way on the coupling constant
of the $\N=4$ theory. Up till now the checks have involved calculating
the effective superpotential in the confining vacuum for the quadratic
$\Tr\,\Phi^2$ deformation \cite{Dijkgraaf:2002dh} and then for
arbitrary polynomial deformations and 
holomorphic condensates in the confining vacuum \cite{us1}. In the
present note we extend this analysis to cover all the massive vacua of
the theory, including the Higgs vacuum. 

We find that the result of the 
$\Tr\,\Phi^2$ deformation is an effective superpotential in each
massive vacuum of the $\N=1^*$ theory computed using the DV
prescription from the matrix model of the form 
\EQ{W_{\rm eff}=-{Np^2\over 12}E_2(p(p\tau+k)/N)\ ,
\qquad{k=0,1,\ldots,N/p-1\ .\label{result}} 
}
where $p$ is a divisor of $N$. Once the vacuum-independent constant
$N^2E_2(\tau)/12$ is added, the results
in different vacua are related to each other by the action of the
modular group, $\SL(2,\mathbb Z)$ on $\tau$, the $\N=4$ coupling, and
the results agree perfectly with
earlier computations using different approaches
\cite{nick,oferandus}. What is remarkable about the derivation
from the matrix model is that the modular properties 
emerge as a result of the computation without being assumed at the beginning.

Let us first briefly recall the vacuum structure of the $\N=1^*$ theory. 
In $\N=1$ language, the $\N=4$ $\SU(N)$ theory with coupling
constant $\tau\equiv 4\pi i/g^2_{YM}+\theta/2\pi$ has three adjoint
chiral superfields $\Phi^+,\Phi^-$ and $\Phi$. We consider a general
class of deformations of the $\N=4$ theory specified by a tree level
superpotential
\EQ{W={1\over
g^2_{YM}}\Tr\left(i\Phi[\Phi^+,\Phi^-]+\Phi^+\Phi^-+V(\Phi)\right)\ ,
}
where $V(\Phi)$ is a general polynomial
\EQ{V(\Phi)\equiv \sum_p g_p\,\Phi^p.}
Note that for the sake of simplicity we have set all masses
to unity in the knowledge that it is simple to re-introduce them. 
For $V(\Phi)=\Phi^2$ we obtain the basic $\N=1^*$
theory, however, it is also interesting to consider the space of possible
$\N=1^*$ deformations as in \cite{us1}. 

The vacuum structure can be deduced by proceeding with a classical
argument augmented by some quantum considerations. The classical
vacuum structure follows by solving the $F$- and $D$-flatness
conditions modulo gauge transformations. Equivalently, one solves the
$F$-flatness conditions modulo complex gauge transformations. Up to
complex gauge transformations, the 
solutions of the $F$-flatness conditions 
are associated to representations of $\SU(2)$ in the
following way: $\Phi$ is precisely $iJ_3$ of an $\SU(2)$
representation, reducible or irreducible, 
of dimension $N$, and
$\Phi^\pm$ have the same non-zero elements as $J^\pm$ except that the
actual numerical values depend on the form of the potential 
$V(\Phi)$. Vacua can be classified
according to whether there are, after the Higgs mechanism, 
unbroken $\U(1)$'s, in which case it is
a massless vacuum. On the contrary, if the unbroken gauge group is
non-abelian (or empty) one has a massive vacuum. In terms
of $\SU(2)$ representations, the massive vacua correspond to $N/p$
copies of the $p$-dimensional representation, where, by construction,
$p$ must be a divisor of $N$. The unbroken gauge group
is then $\SU(N/p)$. The classical eigenvalues of $\Phi$ are explicitly
\EQ{
\lambda_j^{\rm cl}=\tfrac i2(p-2j+1)\ ,\qquad j=1,\ldots,p\ ,
\label{cle}
}
each with a degeneracy of $N/p$.
Now quantum reasoning is needed to deduce the
multiplicity of vacua for each $p$. One
expects at low energies that the theory flows 
to pure $\N=1$ supersymmetric gauge theory with
gauge group $\SU(N/p)$. Standard arguments then imply there are
$N/p$ physically inequivalent quantum vacua. All-in-all, there are
$\sum_{p|N}N/p\equiv\sum_{p|N}p$ massive vacua, where $p$ is a divisor of $N$.

It is the goal of this paper to calculate \eqref{result} from the
matrix model proposal of Dijkgraaf and Vafa
\cite{Dijkgraaf:2002fc,Dijkgraaf:2002vw,talk,Dijkgraaf:2002dh},
generalizing the results of \cite{Dijkgraaf:2002dh,us1}, which held for
the confining vacuum ($p=1$ above).

\section{The matrix model and its solution}

As a direct consequence of the proposal of Dijkgraaf and Vafa
\cite{Dijkgraaf:2002fc,Dijkgraaf:2002vw,talk,Dijkgraaf:2002dh}, in a
given vacuum the exact effective 
superpotential for the above class of deformations of the $\N=4$ theory
is computed by the planar diagram expansion, {\it i.e.\/}~a large-$N$ limit,
of the matrix model for which there is a matrix variable for each
chiral multiplet of the field theory and whose action is the
tree-level superpotential of the field theory. Using the same notation
for the matrix variables and their associated superfields, the matrix model
partition function is  
\EQ{
Z=\int [d\Phi^+][d\Phi^-][d\Phi]\exp-{1\over
g_s}\Tr\left(i\Phi[\Phi^+,\Phi^-]+\Phi^+\Phi^-+V(\Phi)\right)\ 
\label{mm}
} 
In the matrix model, unlike in the field theory, one takes 
$\Phi^+=(\Phi^-)^\dagger$ and the fluctuations of $\Phi$ around the
ensuing saddle points to be Hermitian. At the saddle-point we will
allow the eigenvalues of $\Phi$ to be complex.
As noted for example in \cite{Kazakov:1998ji}, the key fact that permits
the solution of these matrix models is that one may integrate out
$\Phi^\pm$ exactly to obtain a one-matrix integral to be solved in the
large-$N$ limit:
\EQ{Z=\int[d\Phi] {e^{-{1\over g_s}\Tr V(\Phi)}\over{\det(
\text{Adj}_\Phi+i)}}.}
The resulting one-matrix model actually becomes tractable in the large-$N$
limit by going to the eigenvalue basis and performing a large-$N$
saddle-point approximation to the integral. The details of this
procedure have been extensively discussed in the literature. We refer
the reader to \cite{Kazakov:1998ji}, and references therein, for more details. 

In order to describe the confining vacuum, for which $\Phi=0$
classically, one takes a one cut solution as follows 
\cite{Dijkgraaf:2002dh,us1}. The eigenvalues $\lambda_j$ of
$\Phi$ interact via a repulsive effective potential and form a
continuum in the large-$N$ limit and   
condense onto a cut along the real axis. The actual extent of the
cut and the density of eigenvalues $\rho(\lambda)$ along the cut is
self-consistently determined by the saddle-point equation in terms of
the parameters of the deformation $V(\Phi)$ and the matrix model 't~Hooft
coupling $S=g_sN$. Proceeding along these lines leads to a description
of the confining vacuum \cite{Dijkgraaf:2002dh,us1}.

For a general massive vacuum associated in the nomenclature of
$\SU(2)$ representations to $N/p$ copies of the $p$-dimensional
representation, the classical eigenvalues of $\Phi$ are given in \eqref{cle}.
This suggests in the matrix
model we should take a multi-cut solution where at large $N$ 
the eigenvalues condense on $p$ cuts defined
as
\EQ{
{\EuScript C}_j=\Big\{\lambda=\lambda_j^{\rm cl}+x\ ,\quad
-\alpha_j\leq x\leq\alpha_j\Big\}\ ,\quad j=1,\ldots,p\ . 
}

The saddle-point equation in the large-$N$ limit is
most conveniently written in terms of the resolvent function
\EQ{\omega(z)=\int{\rho(\lambda)\over{z-\lambda}}d\lambda
}
where $\rho(\lambda)$ is the unit normalized spectral density of eigenvalues:
\EQ{
\rho(\lambda)=\frac1N\sum_{i=1}^N\delta(\lambda-\lambda_i)\ ,
}
a function that in the large-$N$ limit
only has non-trivial support on the cuts. The
spectral density is normalized in such a way that the filling
fractions of eigenvalues along each cut are given by
\EQ{
\frac{N_j}N=\int_{-\alpha_j}^{\alpha_j}\rho\big(\lambda_j^{\rm cl}+x\big)\,dx\
.
\label{fillf}
}
where $N_j$ denotes the number of eigenvalues along the $j$th cut.

The resolvent $\omega(z)$ is an analytic function on the complex
$z$-plane which has cuts precisely along each ${\EuScript C}_j$, $j=1,\ldots,p$.
The discontinuity across each cut gives the spectral density at that
point:
\EQ{\omega(\lambda+i\epsilon)-\omega(\lambda-i\epsilon)=-2\pi
i\rho(\lambda);
\qquad\lambda\in{\EuScript C}_j\ .}
In terms of $\omega(z)$, the saddle-point equation is
\SP{
\frac{V'(\lambda)}{S}&=\omega(\lambda+i\epsilon)+\omega(\lambda-i\epsilon)\\
&\qquad-
\tfrac12\big(\omega(\lambda+i+i\epsilon)+\omega(\lambda+i-i\epsilon)
+\omega(\lambda-i+i\epsilon)+\omega(\lambda-i-i\epsilon)\big)
\label{spe}
}
for $\lambda\in\cup_j{\EuScript C}_j$.
Notice that various principal
values have been taken, giving rise to the $\pm i\epsilon$ prescriptions, 
in order to render the integrals well defined. Unlike
the single cut solution, the final terms---those shifted by $\pm
i$---also have to be defined as a principal value since most of the shifted
cuts collide with each other in pairs.

At this point we need to employ some extra guesswork. We start with
the observation that in the massive vacuum there should be same number
of eigenvalues on each cut. In other words the filling fractions
\eqref{fillf} should be the same. This is well motivated by the
classical solutions for massive vacua of the {\it field theory} wherein the
adjoint scalars have 
VEVs that split into $N/p$ copies of a $p$-dimensional representation
of $\SU(2)$ preserving an $\SU(N/p)$ gauge symmetry classically.
Of course, we would also expect
solutions of the saddle-point equations with different filling
fractions which would correspond to massless vacua. However, we are
going to specialize to the massive vacua by making an ansatz which
automatically ensures that the filling fractions are identical.
The ansatz we make is that the eigenvalue density is the same along
each of the cuts (all of which have the same extent
$\alpha_j=\alpha$): for each pair $j,k$
\EQ{
\rho\big(\lambda_j^{\rm cl}+x\big)=\rho\big(\lambda_k^{\rm cl}+x\big)\ ,\qquad 
-\alpha\leq x\leq\alpha\ ,
\label{ansatz}
}
or equivalently
\EQ{
\omega(\lambda_j^{\rm cl}+x+i\epsilon)-\omega(\lambda_j^{\rm cl}
+x-i\epsilon)
=\omega(\lambda_k^{\rm cl}+x+i\epsilon)-\omega(\lambda_k^{\rm cl}
+x-i\epsilon)\ .
\label{disc}
}

Having made this ansatz we now investigate the saddle-point equations
\eqref{spe}. Notice that they are linear in the
$w(\lambda+i\epsilon)+w(\lambda-i\epsilon)$ and so we can solve for
these quantities at each cut. One finds 
\SP{
&\omega(\lambda_j^{\rm cl}+x+i\epsilon)+
\omega(\lambda_j^{\rm cl}+x-i\epsilon)\\
&\qquad\qquad=2Y(\lambda_j^{\rm cl}+x)+\tfrac2{p+1}
\big((p+1-j)\omega(x+i(p+1)/2)+j\omega(x-i(p+1)/2)\big)\ , 
\label{nspe}
}
where $-\alpha\leq x\leq\alpha$ and $j=1,\ldots,p$. Notice that the
resolvent has no discontinuities at $x\pm i(p+1)/2$. 
In the above,
$Y(\lambda)$ is the polynomial which is determined by the potential 
$V(\lambda)$ via
\EQ{
\frac{V'(\lambda)}S=2Y(\lambda)-Y(\lambda+i)-Y(\lambda-i)\ .
}
As in the one-cut solution, it useful to define the function 
\EQ{G(z)=U(z)+iS\big(\omega(z+\tfrac i2)-\omega(z-\tfrac i2)\big)
\label{defG}}
where $U(z)$ is a polynomial in $z$ such that
\EQ{iV^\prime(z)=U(z+\tfrac i2)-U(z-\tfrac i2)\ .
\label{defU}}
Given the ansatz \eqref{ansatz}, the analytic structure of $G(z)$ is
much simpler than that of the resolvent itself. The cuts of $\omega(z)$ 
are shifted up and down by $\pm\frac i2$ and because of \eqref{disc}
most cancel pairwise. There are only two remaining cuts---the ones on the
outside of the stack---given by ${i\over 2}p+x$ and $-{i\over 2}p+x$,
respectively, for $-\alpha\leq x\leq\alpha$. Furthermore, 
putting \eqref{nspe} and \eqref{disc} together,
we end up with a rather simple and suggestive constraint on $G(z)$:
\EQ{G(x+\tfrac i2p\pm i\epsilon)=G(x-\tfrac{i}2p\mp
i\epsilon);\qquad -\alpha\leq x\leq\alpha\ .\label{glu}
}

Now we see that having made the ansatz 
we now have what is closely related to the one-cut
solution, $p=1$, of \cite{Dijkgraaf:2002dh,us1}. In terms of the
function $G(z)$, the only difference is that the two cuts of $G(z)$ are
shifted from $\pm i/2$ to $\pm ip/2$. Furthermore, 
just as in the $p=1$ case, the equation
\eqref{glu} has the effect of gluing the top (bottom) of the upper cut to
the bottom (top) of the lower cut making what is effectively a
torus. 
This is very important, since on general grounds we expect each
of the massive vacua to be associated to a particular torus with its
own complex structure \cite{nick,nickprem,us,DS}. 

Fortunately, we  
can now almost lift the analysis of the one-cut solution, with only
slight modifications to account for the new positions of the cuts, and we
refer to \cite{us1} for more details and conventions. 
Firstly, we can uniformize the torus in the $u$-plane by means of the mapping 
\EQ{z(u)=-{p\omega_1\over
\pi}\left[\zeta(u)-{\zeta(\omega_1)\over\omega_1 }u\right]
\equiv -\frac p2{\theta_1^\prime({\pi
u\over2\omega_1}|\ttau)\over\theta_1({\pi u\over2\omega_1}|\ttau)}\ .
\label{defz}}
The torus is defined by the quotient of the complex $u$-plane by the lattice
generated by the two complex numbers $2\omega_1$ and $2\omega_2$. The complex
structure of the torus $\ttau=\omega_2/\omega_1$ is a parameter. In effect we
shall see that we have changed variables from $\alpha$---and hence from
$S$---to the new variable $\ttau$. Notice in \eqref{defz}, 
the multiplicative factor of 
$p$ which ensures that the cuts occur in the
requisite positions. Tightly encircling around the upper (lower) 
cut anti-clockwise in the $z$-plane  
corresponds to $\omega_2+2x\omega_1$ 
(or $-\omega_2+2(1-x)\omega_1$) for $0\leq x\leq 1$
on the $u$-plane. The former, or minus the latter, defines the
$A$-cycle of the torus. In addition, 
the point at infinity in the $z$ plane corresponds to $u=0$.

At this point we find it convenient to choose the simplest $\N=1^*$
deformation obtained by taking a potential $V(\Phi)=\Phi^2$, although
it should be said that more
complicated potentials are just as tractable \cite{us1}. In this case,
Eq.~\eqref{defG} fixes the asymptotic form of $G$ to be
\EQ{
G(z)\ \overset{z\to\infty}=\ z^2+{\cal O}(1/z^2)\ .
\label{asy}
}
Given the fact that $G$ must be single valued on the torus---an
elliptic function---this fixes
its form uniquely. Eq.~\eqref{asy} implies that 
in the $u$ plane $G(z(u))$ must have a double pole at
$u=0$ with a set coefficient. Given that \eqref{defz} implies
\EQ{
u=-\frac{p\omega_1}\pi
\frac1z+\frac{p^3\omega_1^2\zeta(\omega_1)}{\pi^3}\frac{1}{z^3}+\cdots\ ,
}
means that
\EQ{
G(z(u))={p^2\omega_1^2\over \pi^2}
\Big[{\wp}(u)-{2\zeta(\omega_1)\over\omega_1}\Big]\label{gquad}.
}
Once again take note of the factor of $p^2$ which will prove
important. 

We are now ready to implement the Dijkgraaf-Vafa prescription. First
of all, recall that the filling fractions for each of the $p$ original
cuts are given as in Eq.\eqref{fillf}. For our ansatz \eqref{ansatz} 
each of the filling fractions are equal and following the analysis in
\cite{Dijkgraaf:2002dh,us1} we can express them as an integral of
$G(z(u))$ around the $A$-cycle of the torus:
\EQ{
2\pi iS_j=-i\int_AG(z(u)){dz(u)\over du}du={p^3\over
12}{dE_2(\ttau)\over d\ttau}\ ,\qquad S_j\equiv g_sN_j\ .
\label{aint}
}
The factor of $p^3$ arising from the $p^2$ and $p$ factors of $G(z)$
and $z(u)$, respectively, will turn out to be crucial. In the above
$E_2$ is the $2^{\rm nd}$ Eisenstein series (see \cite{us1} for definitions).

The other important ingredient in the Dijkgraaf-Vafa proposal is the quantity
$\partial{\cal F}_0/\partial S_j$ being the variation in the genus zero 
free energy of the matrix model upon transporting an eigenvalue in
from infinity to the $j^{\rm th}$ cut 
to infinity. In the present context where we will take all the filling
fractions to be the same, we only need an expression for the sum
$\sum_{j=1}^p\partial{\cal F}_0/\partial S_j$. This quantity 
is then simply related to an integral of the same form as
\eqref{aint}, but now over the $B$ cycle of the torus
$-\omega_1+(2x-1)\omega_2$, $0\leq x\leq1$. The integral is easily evaluated
following \cite{us1}
\EQ{
\sum_{j=1}^p\frac{\partial{\cal F}_0}{\partial S_j}=-i\int_B G(z(u)){dz(u)\over
du}du=p^3\Big({\ttau\over 12}{dE_2(\ttau)\over d\ttau}-{1\over 12}
E_2(\ttau)\Big)\
.
\label{bint}
}
Again the factors of $p$ are crucial: $G(z)$ and $z(u)$ give $p^2$ and
$p$ respectively.

The effective 
superpotential is obtained by extremizing the following expression
with respect to $S_j$:
\EQ{W_{\rm eff}=\sum_{j=1}^p\Big(
N_j\frac{\partial{\cal F}_0}{\partial S_j}-2\pi i\tau S_j\Big)\ .
\label{DVW}
}
In this expression $N_j$ are the number of eigenvalues of
$\Phi$ which are classically equal to $\lambda_j^{\rm cl}$. It is
important to realize that these are not directly related to the 
$N_j$ in the matrix model (although we use the same notation). The
latter are related to the filling fractions $S_j/S$ via
$N_j=S_j/g_s$. In order to
describe the massive vacua we now take the physical $N_j=N/p$. Since
$N_j$ is an integer, $p$ must be a divisor of $N$.
In this case,
after having made the ansatz \eqref{ansatz}, which solves the
saddle-point equations, each of $S_j$ depends upon only one parameter,
namely $\ttau$, the complex structure of the auxiliary torus in the
$u$-plane. Hence, in implementing the Dijkgraaf-Vafa prescription for
the massive vacua, we
only have to extremize the superpotential with respect to $\ttau$.
Substituting our
expressions \eqref{aint} and \eqref{bint}, we have
\EQ{
W_{\rm eff}=Np^2\Big(\frac{\partial{\cal F}_0}{\partial S}
\Big)_{p=1}-2\pi i\tau p^4 
S_{p=1}\ .
}
where we have written the result in terms terms of the ($p=1$) one-cut
solution quantities in order to highlight the $p$ dependence:
\EQ{
2\pi iS_{p=1}=\frac1{12}\frac{dE_2(\ttau)}{d\ttau}\ ,\qquad
\Big(\frac{\partial{\cal F}_0}{\partial S}\Big)_{p=1}=
{\ttau\over 12}{dE_2(\ttau)\over d\ttau}-{1\over 12}E_2(\ttau)\ .
}
It is easy to see that 
\EQ{\frac{\partial W_{\rm eff}}{\partial\ttau}=0\ \Longrightarrow\ 
\ttau=\frac{p^2\tau}N\ .
}
and substituting back we have
\EQ{
W_{\rm eff}=-{Np^2\over12}E_2(p^2{\tau/N})\ .
\label{ressp}
}
where as stated above $p$ has to be a divisor of $N$. 
Upon restoring the mass scales in the problem this result agrees
precisely, up to a vacuum-independent 
additive constant, with the expressions of
\cite{nick,nickprem}. Note that $\SL(2,{\mathbb Z})$ modular
transformations of $\tau$ relate the superpotentials in different
massive vacua with the understanding that $W_{\rm eff}$ has modular
weight 2 and that one must account for the anomalous modular
transformation of $E_2$ by adding the vacuum independent constant $N^2
E_2(\tau)/12$.\footnote{The freedom to add such a constant is an example of the
more general effect of operator mixing (in this case $\Phi^2$ with the
identity) \cite{oferandus}.}
This is remarkable in that $S$-duality of the underlying
$\N=4$ theory was not assumed
to begin with, but rather emerges from the solution of the matrix model.

We want to emphasize that our method also covers
the Higgs vacuum (and other vacua where $p$ scales with $N$) 
which at first sight is puzzling because in the field
theory all the eigenvalues are---classically at
least---non-degenerate. How can they form continua on $N$ separate
cuts in the matrix model when there are only $N$ eigenvalues? The
resolution of this puzzle is the realization that the $N_j$ in the
field theory are not equated to the $N_j=S_j/g_s$ in the matrix model. The
latter can be large while the former can be set to one in \eqref{DVW}.

Notice that our result \eqref{ressp} only describes one of the $N/p$
massive vacua associated in the original analysis to the
$p$-dimensional representation of $\SU(2)$. The other $N/p-1$ such vacua
are obtained by repeatedly performing the modular transformation 
$\tau\to\tau+1$ a sufficient number of times to give \eqref{result}.

Our multi-cut technology 
can also be employed in the $\N=1^*$ deformation of the
Leigh-Strassler theory for which the confining vacuum result was
quoted in \cite{us1}. The result in the $p^{\rm th}$ massive vacuum is
\EQ{W_{\rm 
eff}= -\frac Np{
\cos(\beta/2)\over 4 
\sin^3(\beta/2)}\;\;{\theta_1(\beta p/2|
p^2\tau/N)\over\theta_1^\prime(\beta p/2|p^2\tau/N)}+
{N\over 4\sin^2(\beta/2)}\ .
}

{\bf Acknowledgments}
S.P.K. would like to acknowledge support from a PPARC Advanced Fellowship.

\end{document}